\documentclass[aps,prb,twocolumn,superscriptaddress,showpacs]{revtex4}
\usepackage{epsfig}
\usepackage[10pt]{moresize}
\usepackage{amssymb}
\usepackage{amsmath}
\usepackage{amscd,rotate}
\usepackage{graphicx}
\usepackage{color}
\usepackage[english]{babel}
\usepackage{mathrsfs}
\usepackage{mhchem}
\usepackage{revsymb}
\usepackage{pifont,bm}
\usepackage{amsfonts}
\usepackage{hyperref}
\begin{document}

\title{Probing Majorana bound states via counting statistics of a single
electron transistor}
\author{Zeng-Zhao Li}
\affiliation{Laboratory for Quantum Optics and Quantum Information, Beijing Computational Science Research Center, Beijing 100094, China}
\affiliation{Department of Applied Physics, Hong Kong Polytechnic University, Hung Hom,
Hong Kong, China}
\author{Chi-Hang Lam}
\email{C.H.Lam@polyu.edu.hk}
\affiliation{Department of Applied Physics, Hong Kong Polytechnic University, Hung Hom,
Hong Kong, China}
\author{J. Q. You}
\email{jqyou@csrc.ac.cn}
\affiliation{Laboratory for Quantum Optics and Quantum Information, Beijing Computational Science Research Center, Beijing 100094, China}


\begin{abstract}
We propose an approach for probing Majorana bound states (MBSs) in a nanowire via counting statistics of a nearby charge detector in the form of a single-electron transistor (SET). We consider the impacts on the counting statistics by both the local coupling between the detector and an adjacent MBS at one end of a nanowire and the nonlocal coupling to the MBS at the other end.
We show that the Fano factor and the skewness of the SET current are minimized for a symmetric SET configuration in the absence of the MBSs or when coupled to a fermionic state. However, the minimum points of operation are shifted appreciably in the presence of the MBSs to asymmetric SET configurations with a higher tunnel rate at the drain than at the source. This feature persists even when varying the nonlocal coupling and the pairing energy between the two MBSs. We expect that these MBS-induced shifts can be measured experimentally with available technologies and can serve as important signatures of the MBSs.
\end{abstract}
\pacs{73.21.-b, 85.35.Gv}
\maketitle

\section{Introduction \label{sec:intro}}

Majorana fermions are particles that are their own antiparticles.
In high-energy physics, neutrino being an elementary particle was
suggested as a Majorana fermion\cite{Wilczek2009}. Experiments aiming to
prove this proposal are still on going. Besides the high-energy context
where they arose, it is believed that Majorana fermions can also emerge as quasiparticles in
condensed-matter systems\cite{Beenakker2011aXiv,Alicea2012}. The
search for Majorana bound states (MBSs) in these systems has attracted much
interest not only due to their exotic properties (e.g., non-Abelian
statistics) but also because they are promising candidates for topological
quantum computation\cite{Kitaev2003AnnPhys,Nayak2008RMP}. Several physical
systems have been suggested to support MBSs, including fractional quantum
Hall states\cite{MooreRead1991,NayakWilczek1996,ReadGreen2000}, chiral
\textit{p}-wave superconductors/superfluids\cite{RiceSigrist1995,ReadGreen2000}, surfaces of three-dimensional (3D)
topological insulators in proximity to an \textit{s}-wave superconductor\cite{FuKane2008PRL},
superfluids in the 3He-B phase\cite{Volovik2003,SilaevVolovik2010JLowTempPhys}, and helical edge modes of 2D
topological insulators in proximity to both a ferromagnet and a superconductor\cite{Akhmerov2008PRL}. More recently, it has been shown that a spin-orbit
coupled semiconducting 2D thin film\cite{Sau2010PRL} or a 1D nanowire\cite%
{Kitaev2001PhysUsp,Sau2010PRB,Lutchyn2010PRL,Oreg2010PRL,Sau2012PRB} with
Zeeman spin splitting, which is in proximity to an \textit{s}-wave
superconductor, can also host MBSs.

Providing experimental evidences for the realization of MBSs is of great importance.
Techniques proposed to detect MBSs
include the analysis of the tunneling spectroscopy\cite{Sengupta2001PRB,LiuBaranger2011PRB,BolechDemler2007PRL,Law2009PRL}, the
verification of the nature of nonlocality\cite%
{Akhmerov2008PRL,Tewari2008PRL} or the observation of the periodic
Majorana-Josephson current\cite{FuKane2009PRL}. In particular, the very
recent observation of a zero-bias peak in the differential conductance
through a semiconductor nanowire in contact with a superconducting electrode
indicated the possible existence of a midgap Majorana state\cite{MourikKouwenhoven2012Science}. Such a zero-bias peak was also observed in
subsequent experiments\cite{DengXu2012arXiv,Rokhinson2012Nature,Das2012arXiv}. However, this
zero-bias peak could be due to the Kondo resonance\cite{LeeAguadoFranceschi2012PRL} and also occur in the presence of either disorders\cite{LiuPotterLawLee2012PRL} or a singlet-doublet quantum phase transition\cite{LeeAguadoFranceschi2014Nnano}, corresponding to ordinary Andreev bound
states rather than MBSs. Moreover, a study of a more realistic model of a nanowire with MBSs further indicates a different origin for this observed zero-bias peak\cite{RainisKlinovajaLoss2013PRB}.  
There are several recent works\cite{GolubAvishai11PRL,Lee13PRB,Pillet13PRB,ChengLutchyn14PRX,Vernek14PRB,LiuChengLutchyn15PRB} developed, for example, to distinguish between the Majorana and Kondo origins of the zero-bias conductance peak, but a definite evidence for the zero-bias anomaly due to MBSs is still missing. 
Therefore, further investigations are needed to convincingly reveal the existence of MBSs.

We will focus on the detection of MBSs which exist in pairs at the two ends of a nanowire.
Most previous studies based on a variety of setups
considered a detector coupled locally to an adjacent MBS at one end of the nanowire only\cite{Akhmerov2008PRL,LiuBaranger2011PRB,XinqiLi2012PRB,Fu2010PRL,Flensberg2011PRL}, as the coupling to the other MBS farther away is neglected. For example, 
a quantum dot coupled to a MBS was studied in Ref.~40. 
The current and the shot noise through the quantum dot were calculated. A characteristic feature in the frequency dependence of the shot noise was proposed as a signature for the MBS. The coupling of a quantum dot to two MBSs at both ends of a nanowire has also been studied\cite{LiuBaranger2011PRB}, but only the conductance was reported.
In this work, we study both the local and nonlocal coupling of a single electron transistor (SET) (consisting of a quantum dot and two electrodes) to two MBSs at both ends of a nanowire.
We calculate the full counting statistics (FCS)\cite{Levitov1993JETP,Levitov1996JMP} of electron transport through the SET.
FCS yields all zero-frequency current correlations at once and provides detailed insights into the nature of charge transfer beyond what is available from  conductance measurements
alone\cite{BlanterButtiker2000PhysRep,Nazarov2003}. Importantly, it has also become an
experimentally accessible technique in recent years\cite{Gustavsson2006PRL,Flindt2009PNAS,Ubbelohde2012}.
%
Using the FCS, we calculate the current, Fano factor and skewness as functions of a tunnel rate ratio of the SET. The calculations are performed for various couplings of the SET island with the MBSs. The results are also compared with those for coupling to a fermionic state instead. We will show in the following that in the absence of the MBSs or when coupled to fermionic states, the Fano factor and the skewness are minimized for a symmetric SET. However, in the presence of the MBSs, the minimum points shift appreciably to occur for an asymmetric SET with a higher tunnel rate at the drain than at the source. We propose that these MBS-induced shifts of the minimum points of the Fano factor and the skewness can be used as signatures for the identification of the MBSs.

\section{The MBS-SET model \label{sec:MBSs-SET model}}

The hybrid system consists of two MBSs and a SET as schematically shown in 
Figure~1. With a conventional \textit{s}-wave superconductor and a
modest magnetic field, the MBSs as electron-hole quasiparticle excitations
have been suggested to exist at the two ends of a semiconductor nanowire with strong
spin-orbit coupling\cite{Oreg2010PRL,Sau2010PRL,Sau2012PRB}.
%
%
The SET consists of a metallic island coupled via tunneling barriers to two
electrodes. The energy levels and the tunneling barriers can be tuned by the gate voltages. By assuming a Zeeman splitting much
larger than the MBS-SET coupling strength, the source-drain bias voltage
across the SET, and the tunneling rates with the source and drain
electrodes, the SET island can be modeled by a single resonant level
occupied by a spin-polarized electron.

\begin{figure}[t]
\includegraphics[width=0.45\textwidth,bbllx=66,bblly=406,bburx=444,bbury=805]{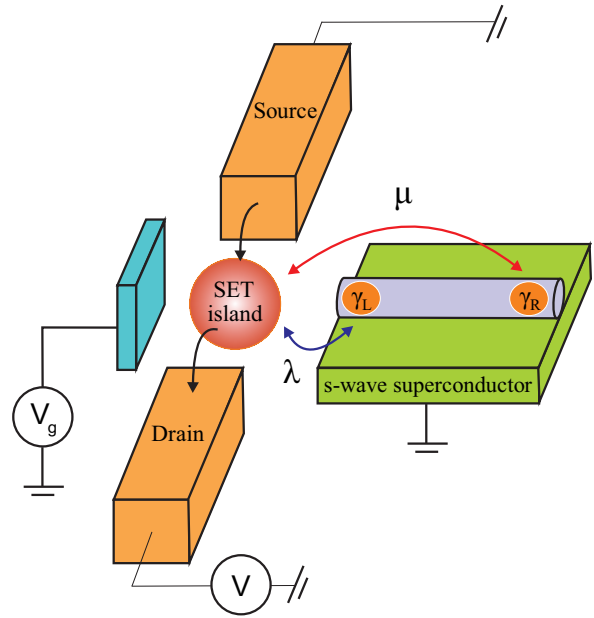} 
\centering
\caption{(Color online) Schematic diagram of the hybrid quantum system
consisting of two MBSs and a SET. The MBSs locate at the two ends of a
nanowire with large Zeeman splitting and strong spin-orbit coupling, which is in proximity to an \textit{s}-wave superconductor. The SET island is coupled to
the source and drain electrodes via tunneling barriers and capacitively
biased by an external gate voltage $V_{g}$. The
energy level of the SET island is tuned to be zero, i.e., in resonance with the MBSs. Also, the SET island couples to the
adjacent MBS with a coupling strength $\lambda$ and the MBS at the other end of the nanowire with a coupling strength $\mu$.}
\label{fig:fig1}
\end{figure}


The interaction between the MBSs and the SET island can be derived from a
second quantization Hamiltonian as (see Methods)
\begin{equation}
H_{t}=\left( d^{\dag }-d\right) \left( \lambda \gamma _{L}+\mu \gamma
_{R}\right) ,  \label{eq:MBSs-SETtunneling}
\end{equation}
where the coupling coefficients $\lambda $ and $\mu $ are assumed to be real
and independent of $k$ for simplicity. This Hamiltonian involves both the
local coupling $\lambda$ to an adjacent MBS at one end of the nanowire and the nonlocal coupling $\mu$
to the MBS at the other end of the nanowire (see 
Figure~1). Due to its smaller magnitude, the nonlocal coupling was neglected in most previous studies
\cite{Akhmerov2008PRL,LiuBaranger2011PRB,BolechDemler2007PRL,Law2009PRL,Tewari2008PRL,XinqiLi2012PRB,Fu2010PRL} with an exception of Ref.~21. 
We note that this nonlocal coupling can give rise to further detector-position-dependent measurement results which may also be used for the identification of the MBSs. The nonlocal coupling is therefore also considered here.

The coupling between two separated MBSs at the two ends of the nanowire can be
described by\cite{Kitaev2001PhysUsp}
\begin{equation}
H_{\gamma }=\frac{i}{2}\varepsilon _{M}\gamma _{L}\gamma _{R},
\end{equation}%
where $\varepsilon_M \sim e^{-l/\zeta}$ is the coupling energy with $l$ being the
wire length and $\zeta$ the superconducting coherent length. The pair of MBSs
can constitute a regular fermion with operators
\begin{equation}
f=\frac{\gamma _{L}+i\gamma _{R}}{2}, \ \ f^{\dag }=\frac{\gamma
_{L}-i\gamma _{R}}{2}.
\end{equation}

In this regular-fermion representation,
the Hamiltonian $H_{\mathrm{sys}}=\varepsilon_I d^{\dag}d+H_{t}+H_{\gamma}$ of the hybrid MBS-SET system becomes
\begin{eqnarray}
H_{\mathrm{sys}} &=&\varepsilon _{I}d^{\dag }d+\varepsilon _{M}\left( f^{\dag }f-%
\frac{1}{2}\right)-\left( \lambda +i\mu \right) \left( d-d^{\dag }\right)
f^{\dag }  \notag \\
&&+\left( \lambda -i\mu \right) \left( d^{\dag }-d\right) f,
\label{syshamiltonian}
\end{eqnarray}%
where $\varepsilon _{I}$ is the resonant-level energy of the SET island and $%
d^{\dag }(d)$ is the corresponding creation (annihilation) operator. Note
that this energy can be tuned by the gate voltage $V_{g}$ to be zero (i.e., $%
\varepsilon _{I}=0$) to ensure resonant tunnelings between the SET island and
the zero-energy MBSs.
The basis states of the system of interest are given by $\left\vert
n_{d}n_{f}\right\rangle ,$ with $n_{d}$ and $n_{f}$ being $0$ and $1,$ i.e.,
$a\equiv\left\vert 00\right\rangle ,b\equiv\left\vert 01\right\rangle
,c\equiv\left\vert 10\right\rangle ,d\equiv\left\vert 11\right\rangle $.
%
To compare the transport behaviors of the SET in the presence of the MBSs with those of a regular fermionic bound state in the nanowire, we also consider the following system Hamiltonian
\begin{eqnarray}
H_{\mathrm{sys}} &=&\epsilon _{I}d^{\dag }d+\epsilon _{M}\left( f^{\dag }f-%
\frac{1}{2}\right) -\left( \lambda +i\mu \right) df^{\dag } \notag \\
&& +\left( \lambda -i\mu \right) d^{\dag }f,
\label{syshamiltonian_compare}
\end{eqnarray}%
which describes the SET when coupled to a regular fermionic
state.

The Hamiltonian for the source and the drain electrodes of the SET is
described by
\begin{equation}
H_{\mathrm{leads}}=\sum_{k}(\omega _{sk}c_{sk}^{\dag }c_{sk}+\omega
_{dk}c_{dk}^{\dag }c_{dk}),
\end{equation}%
where $c_{sk}$ ($c_{dk}$) is the annihilation operator for electrons in the
source (drain) electrode. The tunneling Hamiltonian between the SET island
and the two electrodes is
\begin{equation}
H_{\mathrm{T}}=\sum_{k}[(\Omega _{sk}c_{sk}^{\dag }d+\Omega _{dk}\Upsilon
^{\dag }c_{dk}^{\dag }d)+\mathrm{H.c.}],
\end{equation}%
where $\Omega _{sk(dk)}$ is the coupling strength between the SET island and
the source (drain) electrode. The counting operator $\Upsilon $ ($\Upsilon
^{\dagger }$) decreases (increases) the number of electrons that have
tunneled into the drain electrode in order to keep track of the tunnelings
of successive electrons. Thus, the total Hamiltonian of the system is given by $H_{\mathrm{tot}}=H_{\mathrm{sys}}+H_{\mathrm{leads}}+H_{\mathrm{T}}$.

\section{Counting statistics \label{sec:fcs}}

To study the FCS, it is essential to know the probability $P\left( n,t\right)$ of $n$ electrons having been transported from the SET island to the drain electrode during a period of time interval $t$. It is related to the cumulant generating function $G\left( \chi, t \right) $ defined by\cite{Nazarov2003}
\begin{equation}
e^{-G\left( \chi, t \right) }\!\!=\!\!\sum_{n}P\left( n,t\right) e^{in\chi }.
\label{eq:CGF}
\end{equation}
We will consider the time interval $t$ much longer than the
time for an electron to tunnel through the SET island (i.e., the zero-frequency limit), so that transient properties are insignificant. The derivatives of $G\left( \chi, t \right)$ with respect to the counting field $\chi $ at $\chi =0$ yield the
cumulant of order $m$ as
\begin{equation}
C_{m}=-\left( -i\partial _{\chi }\right) ^{m}G\left( \chi, t \right) |_{\chi
\rightarrow 0}.
\label{eq:CGF-1}
\end{equation}
These cumulants carry complete information of the FCS on the SET island. For
instance, the average current and the shot noise can be expressed as $I=eC_{1}/t$ and $S=2e^{2}C_{2}/t$. Thus, the Fano factor $F$ is given by $F =S/2eI=C_{2}/C_{1}$, which is used to characterize the
bunching and anti-bunching phenomena in the transport process.
The third-order cumulant $C_{3}$ gives rise to the skewness $K=C_{3}/C_{1}$ of the
distribution of transported electrons. 

On the other hand, the probability distribution function of the transported electrons can be expressed as
\begin{equation}
P\left( n,t\right) =\rho _{aa}^{\left(n\right)}\left( t\right) +\rho _{bb}^{\left(n\right)}\left(
t\right) +\rho _{cc}^{\left(n\right)}\left( t\right) +\rho _{dd}^{\left(n\right)}\left( t\right) ,
\label{eq:pdf}
\end{equation}%
where $\rho _{ij}^{\left(n\right)}\left( t\right) $ $(i,j\in \{a,b,c,d\})$ denote the reduced density matrix elements of the SET island at a given number $n$ of electrons being transported from the SET island to the drain electrode at time $t$. We will calculate these reduced density matrix elements using a master equation (see
Methods) which assume a large bias voltage across the SET.
In fact, this large-bias case was considered in many previous studies\cite{NazarovStruben1996PRB,GurvitzPrager1996PRB,Gustavsson2008AdvSolidStatePhys} as it is easy to implement in experiments. Moreover, this makes the problem simpler and more transparent because the broadening effect of the SET level can be neglected (see, e.g., Refs.~53 
and~54
).
Using the discrete Fourier transform of the density
matrix elements given by
\begin{equation}
\rho _{ij}\left( \chi ,t\right) =\sum_{n}\rho _{ij}^{\left(n\right)}\left( t\right)
e^{in\chi },
\end{equation}%
we can convert the master equation into
\begin{equation}
\frac{d}{dt}\varrho =\mathcal{M}(\chi)\varrho ,
\end{equation}%
with
\begin{equation}
\mathcal{M}(\chi)=\left(
\begin{array}{cccc}
A_{11} & 0 & A_{13} & 0 \\
0 & A_{22} & 0 & A_{24} \\
A_{31} & 0 & A_{33} & 0 \\
0 & A_{42} & 0 & A_{44}%
\end{array}%
\right) ,
\end{equation}%
where $\varrho=(\rho _{aa},\rho _{bb},\rho _{cc},\rho _{dd},\varrho_1,\varrho_2)^{T}$ with $\varrho_1=(\mathrm{Re}\left[ \rho _{ab}\right] ,\mathrm{Im}\left[ \rho _{ab}\right] ,\mathrm{Re}%
\left[ \rho _{ac}\right] ,\mathrm{Im}\left[ \rho _{ac}\right] ,\mathrm{Re}%
\left[ \rho _{ad}\right] ,\mathrm{Im}\left[ \rho _{ad}\right]) ,$ and $\varrho_2=(\mathrm{Re}\left[ \rho _{bc}\right] ,\mathrm{Im}\left[ \rho _{bc}\right] ,\mathrm{Re}\left[ \rho _{bd}\right] ,\mathrm{Im}\left[ \rho _{bd}\right] ,\mathrm{Re}%
\left[ \rho _{cd}\right] ,$ $\mathrm{Im}\left[ \rho _{cd}\right] )$, and
\begin{widetext}
\begin{equation}
A_{11}=\left(
\begin{array}{cccc}
-\Gamma _{S} & 0 & \Gamma _{D}e^{i\chi } & 0 \\
0 & -\Gamma _{S} & 0 & \Gamma _{D}e^{i\chi } \\
\Gamma _{S} & 0 & -\Gamma _{D} & 0 \\
0 & \Gamma _{S} & 0 & -\Gamma _{D}%
\end{array}%
\right) ,\; A_{13}=\left(
\begin{array}{cccc}
-2\mu & -2\lambda & 0 & 0 \\
0 & 0 & 2\mu  & -2\lambda  \\
0 & 0 & -2\mu  & 2\lambda  \\
2\mu & 2\lambda & 0 & 0%
\end{array}%
\right) ,
\end{equation}
\begin{equation}
A_{22}=\left(
\begin{array}{cccc}
-\Gamma _{S} & -\varepsilon _{M} & \mu  & -\lambda  \\
\varepsilon _{M} & -\Gamma _{S} & \lambda  & \mu  \\
-\mu  & -\lambda  & -\frac{1}{2}(\Gamma _{S}+\Gamma _{D}) & -\varepsilon _{I}
\\
\lambda  & -\mu  & \varepsilon _{I} & -\frac{1}{2}(\Gamma _{S}+\Gamma _{D})%
\end{array}%
\right) ,\; A_{24}=\left(
\begin{array}{cccc}
-\mu & -\lambda & \Gamma _{D}e^{i\chi } & 0 \\
-\lambda & \mu & 0 & \Gamma _{D}e^{i\chi } \\
0 & 0 & -\mu & -\lambda \\
0 & 0 & -\lambda & \mu
\end{array}%
\right) ,
\end{equation}%
\begin{equation}
A_{31}=\left(
\begin{array}{cccc}
\mu & 0 & 0 & -\mu \\
\lambda & 0 & 0 & -\lambda \\
0 & -\mu  & \mu  & 0 \\
0 & \lambda  & -\lambda  & 0%
\end{array}%
\right) ,\; A_{33}=\left(
\begin{array}{cccc}
-\frac{1}{2}(\Gamma _{S}+\Gamma _{D}) & -\left( \varepsilon _{M}+\varepsilon
_{I}\right)  & 0 & 0 \\
\varepsilon _{M}+\varepsilon _{I} & -\frac{1}{2}(\Gamma _{S}+\Gamma _{D}) & 0 & 0
\\
0 & 0 & -\frac{1}{2}(\Gamma _{S}+\Gamma _{D}) & \varepsilon _{M}-\varepsilon _{I}
\\
0 & 0 & -\left( \varepsilon _{M}-\varepsilon _{I}\right)  & -\frac{1}{2}(\Gamma
_{S}+\Gamma _{D})%
\end{array}%
\right) ,
\end{equation}%
\begin{equation}
A_{42}=\left(
\begin{array}{cccc}
\mu & \lambda & 0 & 0 \\
\lambda & -\mu & 0 & 0 \\
\Gamma _{S} & 0 & \mu & \lambda \\
0 & \Gamma _{S} & \lambda & -\mu
\end{array}%
\right) ,\; A_{44}=\left(
\begin{array}{cccc}
-\frac{1}{2}(\Gamma _{S}+\Gamma _{D}) & -\varepsilon _{I} & \mu  & \lambda  \\
\varepsilon _{I} & -\frac{1}{2}(\Gamma _{S}+\Gamma _{D}) & -\lambda  & \mu  \\
-\mu  & \lambda  & -\Gamma _{D} & -\varepsilon _{M} \\
-\lambda  & -\mu  & \varepsilon _{M} & -\Gamma _{D}%
\end{array}%
\right) .
\end{equation}%
\end{widetext}
Here $\Gamma _{S(D)}$ is the tunneling rate of the electrons
through the barrier between the SET island and the source (drain) electrode
and it is given by
\begin{equation}
\Gamma _{S(D)}=2\pi g_{S(D)}\Omega _{sk(dk)}^{2},
\end{equation}
where $%
g_{i} $ ($i=S$, or $D$) denotes the density of states at the source or drain
electrode and is assumed to be constant over the relevant energy
range.

The formal solution to the dynamical equation of $\varrho (\chi ,t)$ can be readily obtained as $\varrho
(\chi,t)=e^{\mathcal{M}(\chi )t}\varrho (\chi,0)$. The cumulant generating function then reads $G(\chi,t)=-\ln{\rm Tr}\varrho (\chi,t)$. At long time $t$
(i.e., zero-frequency limit), the cumulant generating function is simplified to\cite{BagretsNazarov2003PRB}
\begin{eqnarray}
G(\chi,t)=-\Lambda_{\mathrm{min}}\left(\chi\right)t,
\label{eq:CGF-simplified}
\end{eqnarray}%
where $\Lambda _{\mathrm{min}}(\chi)$ is the minimal eigenvalue of $\mathcal{M}(\chi)$ and satisfies $\Lambda
_{\mathrm{min}}|_{\chi\rightarrow 0}\rightarrow 0$ due to the probability normalization $\sum_{n}P(n,t)=1$.

\section{Signatures of the MBS\lowercase{s} \label{sec:results}}
\subsection{Current}
Below we consider the zero-temperature case for the SET system since related experiments are usually performed at extremely low temperatures (see, e.g., Refs.~54 
and~56
).
Figure~2(a) 
shows the current flowing from the SET island
to the drain electrode as a function of $\Gamma_{D}/\Gamma _{S}$ for $\varepsilon_M=0$ and various values of $\lambda$ and $\mu$. In particular, the case of  $\lambda=\mu=0$ represents the absence of the MBSs. 
Our calculation shows 
that it also equivalently represents the case of coupling to a fermion in the nanowire.
This is expected because a regular fermion state does not affect the counting statistics of a nearby SET in the zero-frequency limit (or stationary behaviors) considered. 
It is
clear from Figure~2(a) that for a symmetric SET in which the tunneling rates between the SET island and the two
electrodes are the same, i.e., $\Gamma_{D}=\Gamma _{S}$, the current does
not vary with $\lambda$ and $\mu$ (see also the analytical result below).
However, when $\Gamma_{D}\neq\Gamma _{S}$, the current in the presence of
the MBSs deviates appreciably from that in the absence of the MBSs,
especially in the region $\Gamma_{D}>\Gamma _{S}$. Moreover, Figure~2(b) shows that
the current also changes, albeit slightly, when varying the coupling energy $\varepsilon_M$
of the two MBSs 
.
From Figures~2(a) and 2(b), although coupling to the MBSs does change the current quantitatively, a distinct qualitative feature is lacking. Thus, it is insufficient to use only the current to show the
existence of the MBSs.

Much of the above numerical results can also be obtained from analytic expressions in some special cases.
For $\varepsilon_M=0$, we obtain from 
equation~(\ref{eq:CGF-simplified}) an analytical result for the current:
\begin{equation}
I=\frac{e\Gamma _{D}\left( 4\xi ^{2}+\Gamma _{S}\Gamma_{\mathrm{tot}%
}\right) }{8\xi ^{2}+\Gamma_{\mathrm{tot}}^{2}},  \label{eq:current}
\end{equation}
where $\Gamma_{\mathrm{tot}}=\Gamma _{S}+\Gamma _{D}$ and $\xi =\sqrt{\lambda ^{2}+\mu ^{2}}$. 
Although this symmetric property of the two couplings $\lambda$ and $\mu$ has been noticed before\cite{LiuBaranger2011PRB}, we emphasize that we apply full counting statistics (including the Fano factor and the skewness as shown below) to reveal signatures of the MBSs, which goes beyond the conductance results reported in Ref.~21. 
When the MBSs are absent, i.e., $\xi\rightarrow 0$, 
equation~(20) 
recovers the well-known result
\begin{equation}
I=\frac{e\Gamma_S\Gamma_D}{\Gamma_{\mathrm{tot}}}.
\label{current_lowbiasTemperature}
\end{equation}
Alternatively, with the MBSs coupled and $\Gamma _{S}=\Gamma _{D}=\Gamma $,
the current is reduced to $I=e\Gamma/2$, independent of the values of $%
\lambda$ and $\mu$.

\begin{figure}[htb]
\centering
  \includegraphics[width=0.45\textwidth,bbllx=3,bblly=2,bburx=591,bbury=823]{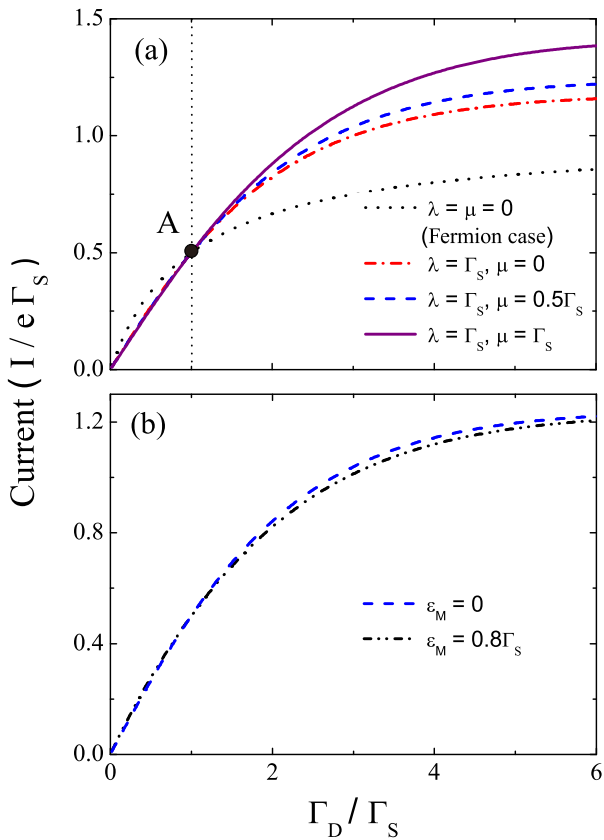}
\caption{(Color online) Current $I$ 
through the SET island to the drain electrode 
versus the
tunneling-rate ratio $\Gamma_{D}/\Gamma_{S}$ for $\varepsilon_I=0$. In (a), $\varepsilon_M=0$ as $\lambda$ and $\mu$ are varied. In (b), $\lambda=\Gamma_S$, and $\mu=0.5\Gamma_S$ as $\varepsilon_M$ is varied.}
\label{fig:fig2}
\end{figure}

When the MBSs are absent, the current through the SET island at zero temperature can also be calculated using\cite{Datta2005}
\begin{equation}
I=\frac{e\Gamma_S\Gamma_D}{\Gamma_{\mathrm{tot}}}\int^{\mu_S}_{\mu_D}dED\left(E\right),
\end{equation}
where $\mu_{S\left( D\right)}$ is the chemical potential of the source (drain) electrode, and $D\left(E\right)$ is the density of states (DOS) of the SET island. When including the electrode-induced level broadening of the SET island, the broadened DOS can be described by a Lorentzian function\cite{Datta2005} centered around $E=\varepsilon_{I}$:
\begin{equation}
D\left(E\right)=\frac{\Gamma_{\mathrm{tot}}/2\pi}{\left(E-\varepsilon_{I}\right)^2+\left(\Gamma_{\mathrm{tot}}/2\right)^2}.
\end{equation}
Therefore, the current can be calculated as
\begin{equation}
I=\frac{e\Gamma_S\Gamma_D}{\Gamma_{\mathrm{tot}}}\mathscr{F}\left(\mu_S,\mu_D\right),
\label{current_Datta}
\end{equation}
where
\begin{equation}
\mathscr{F}\left(\mu_S,\mu_D\right)=\frac{1}{\pi}\arctan\left[\frac{2\Gamma_{\mathrm{tot}}eV}{\Gamma_{\mathrm{tot}}^2+4\left(\varepsilon_{I}-\mu_S\right)\left(\varepsilon_{I}-\mu_D\right)}\right].
\end{equation}
If the bias $eV\equiv\mu_S-\mu_D$ applied on the SET is large
so that the SET level $\varepsilon_I$ is deeply inside the bias window, i.e., $eV\sim 2|\varepsilon_I-\mu_{S\left(D\right)}|\gg \Gamma_{\mathrm{tot}},$ the factor $\mathscr{F}\left(\mu_S,\mu_D\right)$ is simply reduced to
\begin{equation}
\mathscr{F}\left(\mu_S,\mu_D\right)=1.
\end{equation}
Equation~
(\ref{current_Datta}) then recovers\cite{Datta2005} equation~(\ref{current_lowbiasTemperature}).


\subsection{Fano factor}

It is known that the current from the SET island to the drain electrode is
related to the first-order cumulant of the generating function $G(\chi, t)$ by $I=eC_{1}/t$.
The corresponding shot noise is related to the second-order
cumulant of $G(\chi, t)$ as $S=2e^{2}C_{2}/t$. Thus, the Fano factor $F
=S/2eI$ can be written as $F=C_{2}/C_{1}$.
In 
Figure~3(a), we show the Fano factor as a function of $\Gamma_{D}/\Gamma _{S}$ for $\varepsilon_M=0$ and various values of $\lambda$ and $\mu$. 
The black dotted
curve in this figure represents the result not only for the case without the MBSs but also for the identical result for the fermion case similar to that in Figure 1(a).
It
is clear that the Fano factor in the absence of the MBSs reaches its minimum
(i.e., $F_{\mathrm{min}}=1/2$) for a symmetric SET with $\Gamma_{D}/\Gamma_{S}=1$, as
indicated by point B on the black dotted curve in 
Figure~3(a). 
This minimum point of the Fano factor shifts appreciably in the
presence of the MBSs, e.g., $F_{\mathrm{min}}\approx0.49$ at $\Gamma
_{D}/\Gamma_{S}\approx3.58$ when
$\lambda=\Gamma_S$ and $\mu=0$. Interestingly, this shift is robust against varying either the
nonlocal coupling $\mu$ to the more distant MBS or the coupling energy $%
\varepsilon_{M} $ between the two MBSs [see 
Figures~3(a) and~3(b)].

For $\varepsilon_M=0$, an analytical result for the Fano factor can also be
obtained as
\begin{equation}
F=\frac{8\xi ^{2}\left( 8\xi ^{2}+2\Gamma _{S}^{2}-\Gamma
_{D}^{2}+5\Gamma _{S}\Gamma _{D}\right) +\Gamma_{\mathrm{tot}}^{2}\left(
\Gamma _{S}^{2}+\Gamma _{D}^{2}\right) }{\left( 8\xi ^{2}+\Gamma_{\mathrm{%
tot}}^{2}\right) ^{2}}.  \label{eq:fanofactor}
\end{equation}%
Without the MBSs, i.e., $\xi \rightarrow 0,$ we recover the experimentally
verified result\cite{Gustavsson2006PRL} $F=\left( 1+a^{2}\right) /2<1$, where $a=(\Gamma _{S}-\Gamma
_{D})/\Gamma_{\mathrm{tot}}$ is the asymmetry of the SET.
In the presence of
the MBSs and when $\Gamma _{S}=\Gamma _{D}=\Gamma $, the Fano factor follows $%
F=(\Gamma ^{2}+4\xi ^{2})/(2\Gamma ^{2}+4\xi ^{2})$. It depends non-trivially on the couplings between the SET island and the MBSs,
which does not occur for the current (see 
Figure~2). In
Figure~4(a)
, we show the dependence of the minimum
point from equation
~(\ref{eq:fanofactor}) on the SET-MBS coupling. We observe that the
minimum point $(\Gamma_D/\Gamma_S)_{\mathrm{min}}$ of the Fano factor
increases with $\xi/\Gamma_{S}$. This MBS-induced shift of the minimum point of the Fano factor can be used as one of the signatures of the MBSs.
Such a shift does not occur when coupled to a fermion state instead 
[see the black dotted curve in Figure~3(a)]. We emphasize that we have considered a nanowire with both MBSs coupled to the quantum dot\cite{LiuBaranger2011PRB,Flensberg2011PRL}. This generalizes results on the Fano factor from Ref.~40 which considered coupling to only one MBS, which may not be applicable when the distances between the detector (e.g., SET) and two MBSs are comparable.
In addition, instead of considering the Fano factors (or currents) at both the source and drain electrodes as in Ref.~40, 
we find it sufficient to characterize the MBSs by studying  the statistics only at the drain electrode. This is in fact more directly related to typical experimental measurements. In particular, our results on the Fano factor (and also on the skewness as demonstrated below) can reduce back to the experimentally verified ones when the MBSs are decoupled as explained  above.
%
%
Moreover, tuning the gate voltages to control $\Gamma_D/\Gamma_S$ for identifying the MBSs in our proposal is a new alternative to the frequency tuning suggested in Ref.~40. 
%
Note that the shot noise of a quantum dot coupled to a MBS was explored in a more recent work\cite{LiuChengLutchyn15PRB} to distinguish the Majorana origin of the zero-bias anomaly from that due to Kondo effect. However, their results of the shot-noise power spectra as well as the tunneling conductance were obtained under a smaller bias voltage (i.e., $eV\ll\Gamma_{\mathrm{tot}}$). These are quite different from our results of Fano factor (or shot noise) and current which correspond to the case of a large bias voltage (i.e., $eV\gg \Gamma_{\mathrm{tot}}$). In addition, we further explore the skewness below, which goes beyond the differential conductance (or current)\cite{GolubAvishai11PRL,Lee13PRB,Pillet13PRB,ChengLutchyn14PRX,Vernek14PRB,LiuChengLutchyn15PRB} and shot noise\cite{LiuChengLutchyn15PRB} to reveal the signatures of MBSs.


\begin{figure}[t]
\centering
\includegraphics[width=0.45\textwidth,bbllx=3,bblly=2,bburx=590,bbury=806]{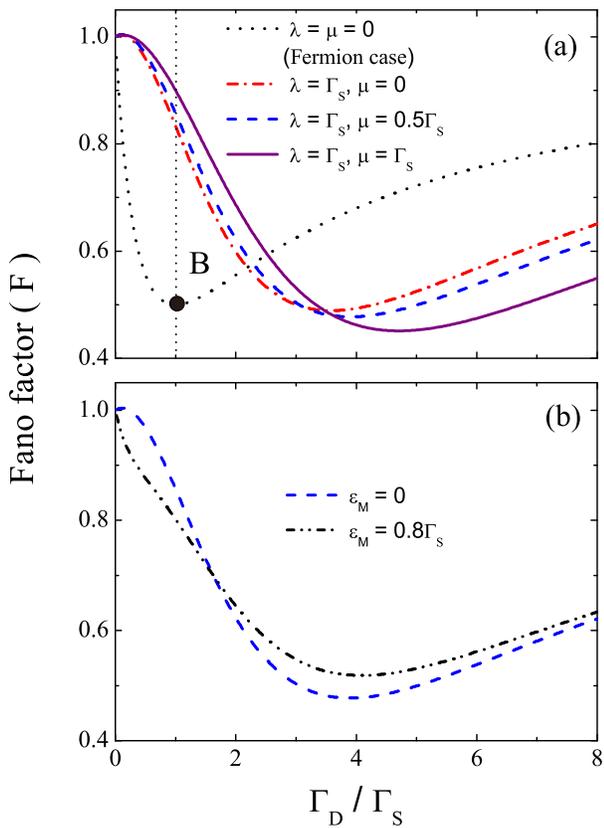}
\caption{(Color online) Fano factor $F$ versus the tunneling-rate ratio $%
\Gamma_{D}/\Gamma_{S}$. The parameters in (a) are the same as those in 
Figure~2(a), and the parameters in (b) are the same as those in Figure~2(b).}
\label{fig:fig3}
\end{figure}

\begin{figure}[htb]
\centering
\includegraphics[width=0.45\textwidth,bbllx=3,bblly=2,bburx=593,bbury=830]{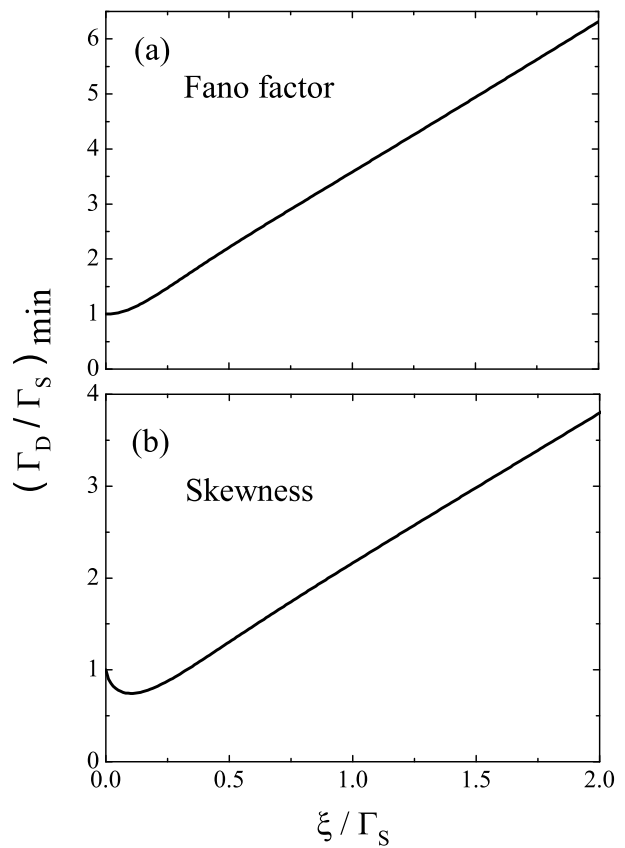}
\caption{The minimum points $(\Gamma_D/\Gamma_S)_{\mathrm{min}}$ of (a) Fano factor and (b) skewness as a function of $\xi=\sqrt{\lambda^2+\mu^2}$.}
\label{fig:fig4}
\end{figure}

\subsection{Skewness}
The skewness of the distribution of transferred electrons is defined as $%
K=C_3/C_1$, which involves the third-order cumulant $C_3$.
Figure~
5(a) shows the skewness for $\varepsilon_M=0$ and various values
of $\lambda$ and $\mu$. It is clear that the skewness in the absence of the MBSs
takes its minimum value (i.e., $K=1/4$) at $\Gamma_D/\Gamma_{S}=1$, as
indicated by point C on the dotted curve. This dotted curve also represents the results for the fermion case 
due to the same reason as that for the result of the current or Fano factor as explained above. 
Also, the minimum point of the
skewness shifts appreciably in the presence of MBSs, e.g., $K_{\mathrm{min}}\approx0.08$ at $\Gamma
_{D}/\Gamma _{S}\approx2.16$ when $\lambda=\Gamma_{S}$ and $\mu=0$.
Moreover, similar to the Fano factor, this shift of the minimum point is also robust against varying $\mu$ and $\varepsilon_M$ [see 
Figures~5(a) and~5(b)].

If $\varepsilon_M=0$, the skewness can be obtained analytically as
\begin{equation}
K=\frac{16(2\xi)^{8}+8(2\xi) ^{6}\mathscr{A}+12(2\xi) ^{4}\mathscr{B}%
+8\xi ^{2}\Gamma_{\mathrm{tot}}^{2}\mathscr{C}+\Gamma_{\mathrm{tot}}^{4}%
\mathscr{D}}{ \left( 8\xi ^{2}+\Gamma_{\mathrm{tot}}^{2}\right) ^{4}},
\label{eq:skewness}
\end{equation}%
%
%
%
where 
\begin{equation}
\mathscr{A}=4\Gamma_{\mathrm{tot}}^{2}+3\Gamma_{D}(\Gamma _{S}-4\Gamma _{D}),
\notag
\end{equation}%
%
%
%
\begin{equation}
\mathscr{B}=\Gamma_{\mathrm{tot}}^{4}-\Gamma_{D}^{2}[7\Gamma_{\mathrm{tot}%
}\Gamma_{S}-(\Gamma _{S}-\Gamma_{D})^{2}],  \notag
\end{equation}%
%
%
%
\begin{equation}
\mathscr{C}=4\Gamma_{\mathrm{tot}}^{4}-3\Gamma_{D}\{3\Gamma_{S}\Gamma_{%
\mathrm{tot}}^{2}+\Gamma
_{D}[3\Gamma_{\mathrm{tot}}^2+2\Gamma_{S}(5\Gamma_{S}-7\Gamma_{D})]\},
\notag
\end{equation}%
%
%
%
\begin{equation}
\mathscr{D}=(\Gamma _{S}-\Gamma_{D})^{4}+2\Gamma _{S}\Gamma _{D}(\Gamma_{%
\mathrm{tot}}^2-2\Gamma _{S}\Gamma_{D}).  \notag
\end{equation}
As expected, when $\xi \rightarrow 0$ (i.e., the case with no MBSs), the
skewness reduces to that of a SET: $K=\mathscr{D}/\Gamma_{\mathrm{tot%
}}^{4}=\left( 1+3a^{4}\right) /4$, which was verified experimentally in Ref.~47
. In the presence of the MBSs,
the skewness takes its minimum value $K_{\mathrm{min}}$ at the minimum point $%
(\Gamma_D/\Gamma_S)_{\mathrm{min}}$ which can be accurately identified from equation
~(\ref{eq:skewness}) and is shown in Figure~
4(b). Similar to the Fano factor, this
MBS-induced shift of the minimum point of the skewness can be used as another
signature of the MBSs.


\begin{figure}[t]
\centering
\includegraphics[width=0.45\textwidth,bbllx=5,bblly=2,bburx=590,bbury=832]{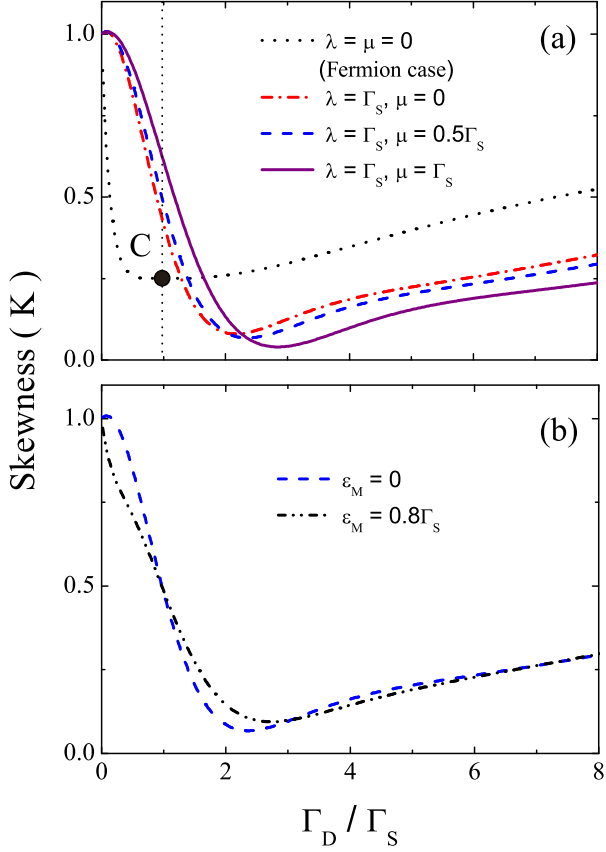}
\caption{(Color online) Skewness $K$ versus the tunneling-rate ratio $%
\Gamma_{D}/\Gamma_{S}$. The parameters in (a) are the same as in 
Figure~2(a), and the parameters in (b) are the same as in Figure~2(b).}
\label{fig:fig5}
\end{figure}

\section{Discussion and Conclusion \label{sec:discussions-conclusions}}

Note that the coupling strengths $\lambda$ and $\mu$ of the SET island to the two MBSs at the ends of the nanowire depend on the position of the detector [see equation
~(\ref{overlapintegral})]. Varying the position of the detector, one can reveal the influence of each MBS on the counting statistics (e.g., the Fano factor and the skewness) of the detector.

In our work, we use the Born-Markov master equation because it is applicable when both the couplings between the system and the electrodes are weak and each electrode has a wide flat energy spectrum. These conditions are valid in our system. Moreover, in studying the counting statistics of the SET island, we need to calculate the $n$-resolved reduced density matrix elements of the SET island [see equation
~(\ref{eq:pdf})]. They are conveniently obtained using the master equation approach.

In summary, we have proposed an experimentally accessible approach to probe
the MBSs via the counting statistics of a charge detector in the form of a
SET. We study the effects of both the local coupling (to an adjacent MBS at one end of the nanowire) and the
nonlocal coupling (to a MBS at the other end of the nanowire) on the counting statistics of the SET
island.
We find that in the presence of the MBSs, the minimum point of both the Fano factor and the skewness shifts appreciably from a symmetric SET configuration to an asymmetric one. This
feature persists even when varying the nonlocal coupling to the farther MBS or
the pairing energy between the two MBSs. These MBS-induced shifts can be used as signatures of the MBSs. Moreover, because our approach
is based on the FCS, it can be readily generalized to higher-order
cumulants to study if they can also be used to probe the MBSs.

\section*{Acknowledgements}

This work is supported by the National Natural Science Foundation of China Grant Nos.~91121015 and 11404019, the National Basic Research Program of China Grant No.~2014CB921401, the NSAF Grant No.~U1330201, Hong Kong GRF Grant No.~501213, and China Postdoctoral Science Foundation Grant No.~2012M520146.

\appendix

\section{Derivation of the tunneling Hamiltonian \label{sec:AppendA}}

For the two MBSs at the ends of a $1$D {\it p}-wave superconductor, which can form at the interface between a semiconductor nanowire with strong spin-orbit coupling and an \textit{s}-wave superconductor\cite{Sau2010PRL,Oreg2010PRL}, the
Majorana operator can be defined as\cite{Oreg2010PRL,Fu2010PRL}
\begin{equation}
\gamma _{i}=\sum_{\sigma }\int dx\left[ f_{i\sigma }^{\ast }\left( x\right)
\psi _{\sigma }\left( x\right) +f_{i\sigma }\left( x\right) \psi _{\sigma
}^{\dag }\left( x\right) \right] ,  \label{MajoranaOperator}
\end{equation}%
where $f_{i\sigma }\left( x\right)$, $i=L$ or $R$, is the Majorana wave function and $\psi
_{\sigma }\left( x\right) $ is the superconductor electron field operator
with spin $\sigma$ $(=\uparrow ,\downarrow )$.

From equation
~(\ref{MajoranaOperator}), it follows that the Majorana operator
satisfies $\gamma^{\dag }_{i}=\gamma_{i} $. The anticommutation relation for the
Majorana operators can be obtained as
\begin{eqnarray}
\left\{ \gamma _{i},\gamma _{j}\right\} &=&\int \int dxdy\big[f_{i\uparrow
}^{\ast }\left( x\right) f_{j\uparrow }\left( y\right) \left\{ \psi
_{\uparrow }\left( x\right) ,\psi _{\uparrow }^{\dag }\left( y\right)
\right\}  \notag \\
&&+f_{i\uparrow }\left( x\right) f_{j\uparrow }^{\ast }\left( y\right)
\left\{ \psi _{\uparrow }^{\dag }\left( x\right) ,\psi _{\uparrow }\left(
y\right) \right\}  \notag \\
&&+f_{i\downarrow }^{\ast }\left( x\right) f_{j\downarrow }\left( y\right)
\left\{ \psi _{\downarrow }\left( x\right) ,\psi _{\downarrow }^{\dag
}\left( y\right) \right\}  \notag \\
&&+f_{i\downarrow }\left( x\right) f_{j\downarrow }^{\ast }\left( y\right)
\left\{ \psi _{\downarrow }^{\dag }\left( x\right) ,\psi _{\downarrow
}\left( y\right) \right\} \big]  \notag \\
&=&\int dx\big[f_{i\uparrow }^{\ast }\left( x\right) f_{j\uparrow }\left(
x\right) +f_{i\uparrow }\left( x\right) f_{j\uparrow }^{\ast }\left( x\right)
\notag \\
&&+f_{i\downarrow }^{\ast }\left( x\right) f_{j\downarrow }\left( x\right)
+f_{i\downarrow }\left( x\right) f_{j\downarrow }^{\ast }\left( x\right) %
\big]  \notag \\
&=&\left\{
\begin{array}{lr}
2&{\rm if }\, i=j, \\
0&{\rm if }\, i\neq j,
\end{array}
\right.  \label{MajoranaAnticommunitator}
\end{eqnarray}%
because of the anticommutation relations for the fermionic field operators%
\begin{eqnarray}
\left\{ \psi _{\alpha }\left( x\right),\psi _{\beta }^{\dag }\left(
y\right) \right\} &=&\delta _{\alpha \beta }\delta \left( x-y\right), \\
\left\{ \psi _{\alpha }\left( x\right) ,\psi _{\beta }\left( y\right)
\right\} &=&0,
\end{eqnarray}%
and the relations of the completeness and orthogonality of the Majorana wave
functions%
\begin{equation}
\int dx\sum_{\sigma }f_{i\sigma }\left( x\right) f_{j\sigma }^{\ast }\left(
x\right) =\delta _{ij}.
\end{equation}%
Obviously, it follows from equation
~(\ref{MajoranaAnticommunitator}) that $\gamma _{i}^{2}=\frac{1}{2}\left\{ \gamma _{i},\gamma _{i}\right\} =1.$

In the Nambu representation of the superconductor electron field operator, $%
\Psi =\left( \psi _{\uparrow },\psi _{\downarrow },\psi _{\downarrow }^{\dag
},\psi _{\uparrow }^{\dag }\right)$. Projecting these field operators onto
the manifold of Majorana states, we have $\Psi \left( x\right)
=\sum_{i}\gamma _{i}$ $\left[ f_{i\uparrow }\left( x\right), f_{i\downarrow
}\left( x\right), f_{i\downarrow }^{\ast }\left( x\right), f_{i\uparrow
}^{\ast }\left( x\right) \right]$.
The electron tunnelings between the MBSs and the SET island are then described by
the Hamiltonian
\begin{eqnarray}
H_{t} &=&\sum_{\sigma }\int dx[t^{\ast }\left( x\right) d_{\sigma }^{\dag
}\psi _{\sigma }\left( x\right)+t\left( x\right) \psi _{\sigma }^{\dag }\left( x\right) d_{\sigma }]
\notag \\
&=&\sum_{i\sigma }\left( V_{i\sigma }^{\ast }d_{\sigma }^{\dag
}-V_{i\sigma }d_{\sigma }\right) \gamma _{i},  \label{overlapintegral}
\end{eqnarray}%
where $V_{i\sigma }^{\ast }=\int dxt^{\ast}\left(x\right) f_{i\sigma
}\left( x\right) ,$ $d_{\sigma }$ is the annihilation operator of the electron with spin $\sigma$ in
the SET island, and $t\left( x\right) $ is the
position-dependent coupling strength between the MBSs and the SET island.
Note that one can always find suitable linear combinations of $d_{\uparrow
}^{\dag }$ and $d_{\downarrow }^{\dag }$ to form spinless fermions $d^{\dag }
$ coupled to the MBSs. Then, the tunneling Hamiltonian becomes
\begin{equation}
H_{t}=\sum_{i}\left( g_{i}^{\ast }d^{\dag }-g_{i}d\right) \gamma _{i},
\label{eq:tunnelingHamiltonian}
\end{equation}%
where operators $d^{\dag }$ and $d$ are defined as%
\begin{equation}
d^{\dag }=\frac{V_{i\uparrow }^{\ast }d_{\uparrow }^{\dag }+V_{i\downarrow
}^{\ast }d_{\downarrow }^{\dag }}{g_{i}^{\ast }}, \ \ d=\frac{V_{i\uparrow }d_{\uparrow }+V_{i\downarrow }d_{\downarrow }}{g_{i}}.
\end{equation}%
If $g_{L}=g_{L}^{\ast }=\lambda $\ and $g_{R}=g_{R}^{\ast }=\mu ,$ we have%
\begin{equation}
H_{t}=\left( d^{\dag }-d\right) \left( \lambda \gamma _{L}+\mu \gamma
_{R}\right) .
\end{equation}%
This is just the Hamiltonian in equation
~(\ref{eq:MBSs-SETtunneling}), which describes the electron tunnelings between the MBSs and the SET island. Note that it includes both the
local coupling $\lambda $\ to the adjacent MBS at one end of the nanowire and the
nonlocal coupling $\mu $ to the MBS at the other end of the nanowire.
Equation
~(\ref{eq:tunnelingHamiltonian}) reduces to the
tunneling Hamiltonian widely used in previous studies (e.g. Refs.~40 
and~41 
) by choosing $\mu=0$.

\section{Quantum dynamics of the SET \label{sec:AppendB}}

Applying the Born-Markov approximation and tracing over the degrees of
freedom of the electrodes coupled to the SET island, the master equation of the hybrid MBS-SET system in the Schr\"{o}dinger picture can be obtained as
\begin{equation}
\dot{\rho}=-i\left[ H_{\mathrm{sys}},\rho \right] +\Gamma _{S}\mathcal{D}\left[
d^{\dag }\right] \rho +\Gamma _{D}\mathcal{D}\left[ \Upsilon _{r}^{\dag }d%
\right] \rho ,  \label{eq:masterEquation}
\end{equation}%
where $\rho \left( t\right) $ is the reduced density operator of the MBS-SET
system, and the superoperator $\mathcal{D},$ acting on any single operator, is
defined as $\mathcal{D}\left[ A\right] \rho =A\rho A^{\dag }-\frac{1}{2}%
A^{\dag }A\rho -\frac{1}{2}\rho A^{\dag }A.$

From equation
~(\ref{eq:masterEquation}) and the relations
\begin{eqnarray}
\langle n|\Upsilon ^{\dagger }\rho \Upsilon |n\rangle &=&\rho
^{(n-1)},~\langle n|\Upsilon \rho \Upsilon ^{\dagger }|n\rangle =\rho
^{(n+1)}, \\
\langle n|\Upsilon ^{\dagger }\Upsilon \rho |n\rangle &=&\rho
^{(n)},~\langle n|\Upsilon \Upsilon ^{\dagger }\rho |n\rangle =\rho ^{(n)},
\end{eqnarray}%
where $n$ is the number of electrons that have tunneled to the drain
electrode, we obtain the equation of motion for each $n$-resolved reduced
density matrix element:
\begin{eqnarray}
\dot{\rho}_{aa}^{\left(n\right)}&=&i\lambda \left( \rho _{ad}^{\left(n\right)}-\rho _{da}^{\left(n\right)}\right) -\mu
\left( \rho _{ad}^{\left(n\right)}+\rho _{da}^{\left(n\right)}\right) -\Gamma _{S}\rho _{aa}^{\left(n\right)} \notag \\
&&+\Gamma_{D}\rho _{cc}^{\left( n-1\right) }, \notag \\
\dot{\rho}_{bb}^{\left(n\right)}&=&i\lambda \left( \rho _{bc}^{\left(n\right)}-\rho _{cb}^{\left(n\right)}\right) +\mu \left(
\rho _{bc}^{\left(n\right)}+\rho _{cb}^{\left(n\right)}\right) -\Gamma _{S}\rho _{bb}^{\left(n\right)}\notag \\
&&+\Gamma _{D}\rho
_{dd}^{\left( n-1\right) },\notag \\
\dot{\rho}_{cc}^{\left(n\right)}&=&-i\lambda \left( \rho _{bc}^{\left(n\right)}-\rho _{cb}^{\left(n\right)}\right) -\mu \left(
\rho _{bc}^{\left(n\right)}+\rho _{cb}^{\left(n\right)}\right) +\Gamma _{S}\rho _{aa}^{\left(n\right)} \notag \\
&&-\Gamma _{D}\rho _{cc}^{\left(n\right)}, \notag \\
\dot{\rho}_{dd}^{\left(n\right)}&=&-i\lambda\left( \rho _{ad}^{\left(n\right)}-\rho _{da}^{\left(n\right)}\right) +\mu
\left( \rho _{ad}^{\left(n\right)}+\rho _{da}^{\left(n\right)}\right) +\Gamma _{S}\rho _{bb}^{\left(n\right)} \notag \\
&&-\Gamma_{D}\rho _{dd}^{\left(n\right)}, \notag \\
\dot{\rho}_{ab}^{\left(n\right)} &=&i\varepsilon _{M}\rho _{ab}^{\left(n\right)}-\left( i\lambda +\mu
\right) \rho _{db}^{\left(n\right)}+\left( i\lambda +\mu \right) \rho _{ac}^{\left(n\right)}-\Gamma
_{S}\rho _{ab}^{\left(n\right)}  \notag \\
&&+\Gamma _{D}\rho _{cd}^{\left( n-1\right) }, \notag \\
\dot{\rho}_{ac}^{\left(n\right)}&=&i\varepsilon _{I}\rho _{ac}^{\left(n\right)}-\left( i\lambda +\mu
\right) \rho _{dc}^{\left(n\right)}+\left( i\lambda -\mu \right) \rho _{ab}^{\left(n\right)} \notag \\
&&-\frac{\Gamma _{S}+\Gamma _{D}}{2}\rho _{ac}^{\left(n\right)}, \\
\dot{\rho}_{ad}^{\left(n\right)} &=&i\left( \varepsilon _{I}+\varepsilon _{M}\right) \rho
_{ad}^{\left(n\right)}+\left( i\lambda +\mu \right) \rho _{aa}^{\left(n\right)}-\left(
i\lambda +\mu \right) \rho _{dd}^{\left(n\right)}  \notag \\
&&-\frac{\Gamma _{S}+\Gamma _{D}}{2}\rho _{ad}^{\left(n\right)}, \notag \\
\dot{\rho}_{bc}^{\left(n\right)} &=&-i\left( \varepsilon _{M}-\varepsilon _{I}\right) \rho
_{bc}^{\left(n\right)}-\left( i\lambda -\mu \right) \rho _{cc}^{\left(n\right)}+\left( i\lambda -\mu \right)
\rho _{bb}^{\left(n\right)}  \notag \\
&&-\frac{\Gamma _{S}+\Gamma _{D}}{2}\rho _{bc}^{\left(n\right)}, \notag \\
\dot{\rho}_{bd}^{\left(n\right)}&=&i\varepsilon _{I}\rho _{bd}^{\left(n\right)}-\left( i\lambda -\mu \right) \rho
_{cd}^{\left(n\right)}+\left( i\lambda +\mu \right) \rho _{ba}^{\left(n\right)} \notag \\
&&-\frac{\Gamma _{S}+\Gamma _{D}}{2}\rho _{bd}^{\left(n\right)}, \notag \\
\dot{\rho}_{cd}^{\left(n\right)}&=&i\varepsilon _{M}\rho _{cd}^{\left(n\right)}-\left( i\lambda +\mu \right) \rho
_{bd}^{\left(n\right)}+\left( i\lambda +\mu \right) \rho _{ca}^{\left(n\right)}+\Gamma
_{S}\rho _{ab}^{\left(n\right)} \notag \\
&&-\Gamma _{D}\rho _{cd}^{\left(n\right)}. \notag
\end{eqnarray}
With the $n$-resolved matrix elements obtained, the reduced density matrix elements are given by $\rho_{ij}=\langle i|\rho|j\rangle=\sum_{n}\rho_{ij}^{\left(n\right)},$ $i,j\in \{a,b,c,d\}$.


\begin{thebibliography}{99}

\bibitem{Wilczek2009} Wilczek, F. Majorana returns. \textit{Nat. Phys.} \textbf{5}, 614-618 (2009).

\bibitem{Beenakker2011aXiv} Beenakker, C. W. J.  Search for Majorana fermions in superconductors. \textit{Annu. Rev. Cond. Mat. Phys.} \textbf{4}, 113-136 (2013).

\bibitem{Alicea2012} Alicea, J. New directions in the pursuit of Majorana fermions in solid state systems. \textit{Rep. Prog. Phys.} \textbf{75}, 076501 (2012).

\bibitem{Kitaev2003AnnPhys} Kitaev, A. Fault-tolerant quantum computation by anyons. \textit{Ann. Phys.} \textbf{303}, 2-30 (2003).

\bibitem{Nayak2008RMP} Nayak, C., Simon, S. H., Stern, A., Freedman, M. \& Sarma, S. D. Non-Abelian anyons and topological quantum computation. \textit{Rev. Mod. Phys.} \textbf{80}, 1083-1159 (2008).

\bibitem{MooreRead1991} Moore, G. \& Read, N. Nonabelions in the fractional quantum hall effect. \textit{Nucl. Phys. B} \textbf{360}, 362-396 (1991).

\bibitem{NayakWilczek1996} Nayak, C. \& Wilczek, F. $2n$-quasihole states realize $2^{n-1}$-dimensional spinor braiding statistics in paired quantum Hall states. \textit{Nucl. Phys. B} \textbf{479}, 529-553 (1996).

\bibitem{ReadGreen2000} Read, N. \& Green, D. Paired states of fermions in two dimensions with breaking of parity and time-reversal symmetries and the fractional quantum Hall effect. \textit{Phys. Rev. B} \textbf{61}, 10267-10297 (2000).

\bibitem{RiceSigrist1995} Rice, T. M. \& Sigrist, M. $\ce{Sr_2RuO_4}$: An electronic analogue of $^3\ce{He}$?  \textit{J. Phys. Cond. Matt.} \textbf{7}, L643-L648 (1995).

\bibitem{FuKane2008PRL} Fu, L. \& Kane, C. L. Superconducting proximity effect and Majorana fermions at the surface of a topological insulator. \textit{Phys. Rev. Lett.} \textbf{100}, 096407 (2008).
%

\bibitem{Volovik2003} Volovik, G. E.  \textit{The Universe in a Helium Droplet} (Oxford University Press, Oxford, 2003).

\bibitem{SilaevVolovik2010JLowTempPhys} Silaev, M. A.  \& Volovik, G. E. Topological superfluid $^3\ce{He}$-$
\ce{B}$: Fermion zero modes on interfaces and in the vortex core. \textit{J. Low. Temp. Phys.} \textbf{161}, 460-473 (2010).

\bibitem{Akhmerov2008PRL} Nilsson, J., Akhmerov, A. R. \& Beenakker, C.W. J. Splitting of a Cooper pair by a pair of Majorana bound states. \textit{Phys. Rev. Lett.} \textbf{101}, 120403 (2008).

\bibitem{Sau2010PRL} Sau, J. D., Lutchyn,  R. M., Tewari, S. \& Sarma, S. D. Generic new platform for topological quantum computation using semiconductor heterostructures. \textit{Phys. Rev. Lett.} \textbf{104}, 040502 (2010).

\bibitem{Kitaev2001PhysUsp} Kitaev, A. Unpaired Majorana fermions in quantum wires. \textit{Phys. Usp.} \textbf{44}, 131-136 (2001).

\bibitem{Sau2010PRB} Sau, J. D., Tewari, S., Lutchyn, R. M., Stanescu, T. \& Sarma, S. D. Non-Abelian quantum order in spin-orbit-coupled semiconductors: Search for topological Majorana particles in solid-state systems. \textit{Phys. Rev. B} \textbf{82}, 214509 (2010).

\bibitem{Lutchyn2010PRL} Lutchyn, R. M., Sau, J. D. \& Sarma, S. D. Majorana fermions and a topological phase transition in semiconductor-superconductor heterostructures. \textit{Phys. Rev. Lett.} \textbf{105}, 077001 (2010).

\bibitem{Oreg2010PRL} Oreg, Y., Refael, G. \& Oppen, F. von. Helical liquids and Majorana bound states in quantum wires. \textit{Phys. Rev. Lett.} \textbf{105}, 177002 (2010).

\bibitem{Sau2012PRB} Sau, J. D., Tewari, S. \& Sarma, S. D. Experimental and materials considerations for the topological superconducting state in electron- and hole-doped semiconductors: Searching for non-Abelian Majorana modes in 1D nanowires and 2D heterostructures. \textit{Phys. Rev. B} \textbf{85}, 064512 (2012).

\bibitem{Sengupta2001PRB} Sengupta, K., Zutic, I., Kwon, H. J., Yakovenko, V. M. \& Sarma, S. D. Midgap edge states and pairing symmetry of quasi-one-dimensional organic superconductors. \textit{Phys. Rev. B} \textbf{63}, 144531 (2001).

\bibitem{LiuBaranger2011PRB} Liu, D. E. \& Baranger, H. U. Detecting a Majorana-fermion zero mode using a quantum dot. \textit{Phys. Rev. B} \textbf{84}, 201308 (2011).

\bibitem{BolechDemler2007PRL} Bolech, C. J. \& Demler, E. Observing Majorana bound states in $p$-wave superconductors using noise measurements in tunneling experiments. \textit{Phys. Rev. Lett.} \textbf{98}, 237002 (2007).

\bibitem{Law2009PRL} Law, K. T., Lee, P. A. \& Ng, T. K. Majorana fermion induced resonant Andreev reflection. \textit{Phys. Rev. Lett.} \textbf{103}, 237001 (2009).

\bibitem{Tewari2008PRL} Tewari, S., Zhang, C., Sarma, S. D., Nayak, C. \& Lee, D. H. Testable signatures of quantum nonlocality in a two-dimensional chiral $p$-wave superconductor. \textit{Phys. Rev. Lett.} \textbf{100}, 027001 (2008).

\bibitem{FuKane2009PRL} Fu, L. \& Kane, C. L. Josephson current and noise at a superconductor/quantum-spin-Hall-insulator/superconductor junction. \textit{Phys. Rev. B} \textbf{79}, 161408 (2009).

\bibitem{MourikKouwenhoven2012Science} Mourik, V., Zuo, K., Frolov, S. M., Plissard, S. R., Bakkers, E. P. A. M. \& Kouwenhoven, L. P. Signatures of Majorana fermions in hybrid superconductor-semiconductor nanowire devices.  \textit{Science} \textbf{336}, 1003-1007 (2012).
%
%

\bibitem{DengXu2012arXiv} Deng, M. T., Yu,  C. L., Huang, G. Y., Larsson,  M., Caroff, P. \& Xu, H. Q. Anomalous zero-bias conductance peak in a $\ce{Nb}$-$\ce{InSb}$ nanowire-$\ce{Nb}$ hybrid device. \textit{Nano Lett.} \textbf{12}, 6414-6419 (2012).

\bibitem{Rokhinson2012Nature} Rokhinson, L. P., Liu, X. \& Furdyna, J. K. The fractional a.c. Josephson effect in a semiconductor-superconductor nanowire as a signature of Majorana particles. \textit{Nat. Phys.} \textbf{8}, 795-799 (2012).

\bibitem{Das2012arXiv} Das, A., Ronen, Y., Most, Y., Oreg,  Y., Heiblum, M. \& Shtrikman, H. Zero-bias peaks and splitting in an $\ce{Al}$-$\ce{InAs}$ nanowire topological superconductor as a signature of Majorana fermions. \textit{Nat. Phys.} \textbf{8}, 887-895 (2012). 

\bibitem{LeeAguadoFranceschi2012PRL} Lee, E. J. H., Jiang, X., Aguado, R., Katsaros, G., Lieber, C. M. \& Franceschi, S. D. Zero-bias anomaly in a nanowire quantum dot coupled to superconductors. \textit{Phys. Rev. Lett.} \textbf{109}, 186802 (2012).

\bibitem{LiuPotterLawLee2012PRL} Liu, J., Potter, A. C., Law, K. T. \& Lee, P. A. Zero-bias peaks in the tunneling conductance of spin-orbit-coupled superconducting wires with and without Majorana end-states. \textit{Phys. Rev. Lett.} \textbf{109}, 267002 (2012).

\bibitem{LeeAguadoFranceschi2014Nnano} Lee, E. J. H., Jiang, X., Houzet, M., Aguado, R., Lieber, C. M. \& Franceschi, S. D. Spin-resolved Andreev levels and parity crossings in hybrid superconductor-semiconductor nanostructures. \textit{Nat. Nanotech.} \textbf{9}, 79-84 (2014).

\bibitem{RainisKlinovajaLoss2013PRB} Rainis, D., Trifunovic, L., Klinovaja, J. \& Loss, D. Towards a realistic transport modeling in a superconducting nanowire with Majorana fermions. \textit{Phys. Rev. B} \textbf{87}, 024515 (2013).

\bibitem{GolubAvishai11PRL} Golub, A., Kuzmenko, I. \& Avishai, Y. Kondo correlations and Majorana bound states in a metal to quantum-dot to topological-superconductor junction. \textit{Phys. Rev. Lett.} \textbf{107}, 176802 (2011).

\bibitem{Lee13PRB} Lee, M., Lim, J. S. \& L\'{o}pez, R. Kondo effect in a quantum dot side-coupled to a topological superconductor. \textit{Phys. Rev. B} \textbf{87}, 241402 (2013).

\bibitem{Pillet13PRB} Pillet, J. D., Joyez, P., \v{Z}itko, R. \& Goffman, M. F. Tunneling spectroscopy of a single quantum dot coupled to a superconductor: From Kondo ridge to Andreev bound states. \textit{Phys. Rev. B} \textbf{87}, 045101 (2013).

\bibitem{ChengLutchyn14PRX} Cheng, M., Becker, M., Bauer, B. \& Lutchyn, R. M. Interplay between Kondo and Majorana interactions in quantum dots. \textit{Phys. Rev. X} \textbf{4}, 031051 (2014).

\bibitem{Vernek14PRB} Vernek, E.,  Penteado, P. H., Seridonio, A. C. \& Egues, J. C. Subtle leakage of a Majorana mode into a quantum dot. \textit{Phys. Rev. B} \textbf{89}, 165314 (2014).

\bibitem{LiuChengLutchyn15PRB} Liu, D. E., Cheng, M. \& Lutchyn, R. M. Probing Majorana physics in quantum-dot shot-noise experiments. \textit{Phys. Rev. B} \textbf{91}, 081405 (2015).

\bibitem{XinqiLi2012PRB} Cao, Y., Wang, P., Xiong, G., Gong,  M. \& Li, X. Q. Probing the existence and dynamics of Majorana fermion via transport through a quantum dot. \textit{Phys. Rev. B} \textbf{86}, 115311 (2012).
%

\bibitem{Fu2010PRL} Fu, L. Electron teleportation via Majorana bound states in a mesoscopic superconductor.
\textit{Phys. Rev. Lett.} \textbf{104}, 056402 (2010).

\bibitem{Flensberg2011PRL} Flensberg, K. Non-Abelian operations on Majorana fermions via single-charge control. \textit{Phys. Rev. Lett.} \textbf{106}, 090503 (2011).

\bibitem{Levitov1993JETP} Levitov, L. S. \& Lesovik, G. B. Charge distribution in quantum shot noise. \textit{JETP Lett.} \textbf{\ 58}, 230-235 (1993).

\bibitem{Levitov1996JMP} Levitov, L. S., Lee, H. W. \& Lesovik, G. B. Electron counting statistics and coherent states of electric current. \textit{J. Math. Phys.} \textbf{37}, 4845-4866 (1996).

\bibitem{BlanterButtiker2000PhysRep} Blanter, Y. \& B\"{u}ttiker, M. Shot noise in mesoscopic conductors. \textit{Phys. Rep.} \textbf{336}, 1-166 (2000).

\bibitem{Nazarov2003} Nazarov, Y. V. \textit{Quantum Noise in Mesoscopic Physics} (Kluwer, Dordrecht, 2003).

\bibitem{Gustavsson2006PRL} Gustavsson, S. \textit{et al.} Counting statistics of single electron transport in a quantum dot. \textit{Phys. Rev. Lett.} \textbf{96}, 076605 (2006).

\bibitem{Flindt2009PNAS} Flindt, C. \textit{et al.} Universal oscillations in counting statistics. \textit{Proc. Natl. Acad. Sci. USA} \textbf{106}, 10116-10119 (2009).

\bibitem{Ubbelohde2012} Ubbelohde, N., Fricke, C., Flindt, C., Hohls, F. \& Haug, R. J. Measurement of finite-frequency current statistics in a single electron transistor. \textit{Nat. Commun.} \textbf{3}, 612 (2012).

\bibitem{Gustavsson2008AdvSolidStatePhys} Gustavsson, S. \textit{et al.} Counting statistics of single electron transport in a semiconductor quantum dot. \textit{Adv. in Solid State Phys.} \textbf{46}, 31-43 (2008).

\bibitem{NazarovStruben1996PRB} Nazarov, Y. V. \& Struben, J. J. R. Universal excess noise in resonant tunneling via strongly localized states. \textit{Phys. Rev. B} \textbf{53}, 15466-15468 (1996).

\bibitem{GurvitzPrager1996PRB} Gurvitz, S. A. \& Prager, Y. S. Microscopic derivation of rate equations for quantum transport. \textit{\ Phys. Rev. B} \textbf{53}, 15932-15943 (1996).

\bibitem{Datta2005} Datta, S. \textit{Quantum Transport: Atom to Transistor} (Cambridge University Press, Cambridge, 2005).

\bibitem{Gustavsson2007PRL} Gustavsson, S. \textit{et al.} Frequency-selective single-photon detection using a double quantum dot. \textit{Phys. Rev. Lett.} \textbf{99}, 206804 (2007).

\bibitem{BagretsNazarov2003PRB} Bagrets, D. A. \& Nazarov, Y. V. Full counting statistics of charge transfer in Coulomb blockade systems. \textit{Phys. Rev. B} \textbf{67}, 085316 (2003).

\bibitem{Kiessich2007PRL} Kie{\ss}lich, G., Sch\"{o}ll, E., Brandes, T., Hohls, F. \& Haug, R. J. Noise enhancement due to quantum coherence in coupled quantum dots. \textit{Phys. Rev. Lett.} \textbf{99}, 206602 (2007).
\end{thebibliography}
\end{document}